\documentclass[a4paper,12pt, epsfig]{article}
\def\letter{0}\def\pr{0}
\usepackage{epsfig}
\usepackage{epstopdf}
\usepackage{graphicx}
\usepackage{ifthen}

\usepackage{amsmath}
\usepackage[psamsfonts]{amssymb}
\usepackage{euscript}

\usepackage{latexsym}
\usepackage[arrow,matrix,curve]{xy}

\newskip\humongous \humongous=0pt plus 1000pt minus 1000pt

\newif\ifdtup

\def\,{\hspace{-.1cm}}
\def\hsp{,\hspace{.7cm}}

\def\ff{{\mathcal{F}}}

\def\fc#1#2 {\frac{n}{q}#1\frac{n}{q}#2}

\def\fc#1#2 {\frac{n}{q}#1\frac{n}{q}#2}

\newcommand{\vac}{\ensuremath{|0\rangle}}

\renewcommand{\cos}{\textrm{cos}}
\renewcommand{\sin}{\textrm{sin}}

\renewcommand{\tanh}{\textrm{tanh}}
\newcommand{\sech}{\textrm{sech}}
\newcommand{\csch}{\textrm{csch}}

\renewcommand{\theequation}{\arabic{section}.\arabic{equation}}
\renewcommand{\(}{\begin{equation}}
\renewcommand{\)}{end{equation} \vspace{-.05in}\linebreak}

\newcounter{saveeqn}
\newcounter{savealpheqn}

\newcommand{\alpheqn}{\setcounter{saveeqn}{\value{equation}}%
  \stepcounter{saveeqn}\setcounter{equation}{0}%
  \renewcommand{\theequation}{\mbox{\arabic{section}.\arabic{saveeqn}
\alph{equation}}}
  \renewcommand{\)}{\end{equation}}}
\def\part#1{\frac{\partial}{\partial{#1}}}%
\def\group#1{\refstepcounter{equation}\setcounter{saveeqn}
 {\value{equation}}%
  \label{#1}\setcounter{equation}{0}%
\renewcommand{\theequation}{\mbox{\arabic{section}.\arabic{saveeqn}
\alph{equation}}}
  \renewcommand{\)}{\end{equation}}}
\newcommand{\reseteqn}{\setcounter{equation}{\value{saveeqn}}%
  \renewcommand{\theequation}{\arabic{section}.\arabic{equation}}%
  \renewcommand{\)}{\end{equation}}}

\newcommand{\aalpheqn}{\setcounter{saveeqn}{\value{equation}}%
  \stepcounter{saveeqn}\setcounter{equation}{0}%
  \renewcommand{\theequation}{\mbox{
        \Alph{subsection}.\arabic{saveeqn}\alph{equation}}}
   \renewcommand{\)}{\end{equation}}}
\newcommand{\areseteqn}{\setcounter{equation}{\value{saveeqn}}%
  \renewcommand{\theequation}{\Alph{subsection}.\arabic{equation}}%
  \renewcommand{\)}{\end{equation}}}

\renewcommand{\thefootnote}{\alph{footnote}}
\renewcommand{\(}{\begin{equation}}
\renewcommand{\)}{\end{equation}}
\newcommand{\ba}{\begin{eqnarray}}
\newcommand{\ea}{\end{eqnarray}}
\renewcommand{\a}{\alpha}
\renewcommand{\b}{\beta}

\newcommand{\cbp}{\mathop{\vtop{\ialign{##\crcr
   $\hfil\displaystyle{}\hfil$\crcr\noalign{\kern-13pt\nointerlineskip}
   \BIG{)}\hskip0pt\crcr\noalign{\kern3pt}}}}}
\newcommand{\pa}{\mathop{\vtop{\ialign{##\crcr

$\hfil\displaystyle{\oplus}\hfil$\crcr\noalign{\kern+1pt\nointerlineskip
}
   \hspace{.08in}$^{\alpha=0}$\hskip6pt\crcr\noalign{\kern3pt}}}}}
\renewcommand{\hsp}{,\hspace{.3in}}
\newcommand{\p}{^\prime}
\newcommand{\pp}{^{\prime\prime}}
\newcommand{\w}{\omega}

\def\H{\ensuremath{\ES{H}}}
\def\i{\ensuremath{\dot\imath}}

\def\D{\ensuremath{{\cal D}}}

\catcode`\@=11
\def\vereq#1#2{\lower3pt\vbox{\baselineskip1.5pt \lineskip1.5pt
\ialign{$\m@th#1\hfill##\hfil$\crcr#2\crcr\sim\crcr}}}
\catcode`\@=12

\renewcommand{\(}{\begin{equation}}
\renewcommand{\)}{\end{equation}}

\def\pin#1{\int \frac{d#1}{2\pi}}

\def\Bd#1{B^\dag_{k_{#1}}}

\def\Bk#1{B_{k_{#1}}}

\def\tp#1#2#3{\hbox{\rm tan}^#1(\thet#2#3)}

\def\df{\mathcal{D}_f}

\def\I{\mathcal{I}}

\def\Bd1k{B^\dag_k}
\def\B1k{B_k}

\def\w#1{\omega_{k_{#1}}}

\def\pin#1{\int \frac{d#1}{2\pi}}
\def\ppin#1{\int\hspace{-17pt}\sum \frac{d#1}{2\pi}}
\def\ppink#1{\int\hspace{-17pt}\sum\frac{d^{#1}k}{(2\pi)^{#1}}}
\def\ppinkp#1{\int\hspace{-17pt}\sum\frac{d^{#1}k\p}{(2\pi)^{#1}}}

\def\Bd#1{B^\dag_{k_{#1}}}

\def\tp#1#2#3{\hbox{\rm tan}^#1(\thet#2#3)}

\def\cc{\mathcal{C}}
\def\df{\mathcal{D}_{f}}

\def\B#1{B^\dag_{k_{#1}}}

\def\I{\mathcal{I}}

\def\os{\omega_S}

\def\as{|\alpha;\sigma\rangle}

\newcommand{\g}{{\mathfrak g}}

\def\ch{{\mathcal{H}}}

\def\tp{{\tilde{\phi}}}
\def\ok#1{\omega_{k_{#1}}}

\def\V#1{V^{(#1)}(gf(x))}

\def\o{\omega_k}

\newcommand{\bal}{\begin{aligned}}
\newcommand{\eal}{\end{aligned}}

\newcommand{\beas}{\begin{eqnarray*}}
\newcommand{\eeas}{\end{eqnarray*}}

\newcommand{\bquo}{\begin{quote}}
\newcommand{\enqu}{\end{quote}}

\def\ch{{\mathcal{H}}}

\def\ok#1{\omega_{k_{#1}}}

\def\V#1{V^{(#1)}[gf(x)]}

\def\os{\omega_S}

\def\as{|\alpha;\sigma\rangle}

\def\H{\mathcal{{H}}}

\newcommand{\beq}{\begin{equation}}
\newcommand{\eeq}{\end{equation}}
\newcommand{\bea}{\begin{eqnarray}}
\newcommand{\eea}{\end{eqnarray}}

\newskip\humongous \humongous=0pt plus 1000pt minus 1000pt

\newif\ifdtup

\catcode`\@=11

\ifthenelse{\equal{\letter}{0}}{ 

\@addtoreset{equation}{section}
\def\theequation{\arabic{section}.\arabic{equation}}

\def\@normalsize{\@setsize\normalsize{15pt}\xiipt\@xiipt
\abovedisplayskip 14pt plus3pt minus3pt%
\belowdisplayskip \abovedisplayskip
\abovedisplayshortskip \z@ plus3pt%
\belowdisplayshortskip 7pt plus3.5pt minus0pt}

\def\small{\@setsize\small{13.6pt}\xipt\@xipt
\abovedisplayskip 13pt plus3pt minus3pt%
\belowdisplayskip \abovedisplayskip
\abovedisplayshortskip \z@ plus3pt%
\belowdisplayshortskip 7pt plus3.5pt minus0pt
\def\@listi{\parsep 4.5pt plus 2pt minus 1pt
      \itemsep \parsep
      \topsep 9pt plus 3pt minus 3pt}}

\relax

\def\section{\@startsection{section}{1}{\z@}{3.5ex plus 1ex minus  .2ex}{2.3ex plus .2ex}{\large\bf}}

\def\thesection{\arabic{section}}
\def\thesubsection{\arabic{section}.\arabic{subsection}}

\def\appendix{\setcounter{section}{0}
 \def\thesection{Appendix \Alph{section}}
 \def\thesubsection{\Alph{section}.\arabic{subsection}}
 \def\theequation{\Alph{section}.\arabic{equation}}}
\renewcommand{\theequation}{\arabic{section}.\arabic{equation}}

}{
\renewcommand{\theequation}{\arabic{equation}}

} 

\begin{document}
\def\thefootnote{\fnsymbol{footnote}}
\def\thetitle{Leading Quantum Correction to the $\Phi^4$ Kink Form Factor }
\def\autone{Hengyuan Guo}
\def\affa{Institute of Modern Physics, NanChangLu 509, Lanzhou 730000, China}
\def\affb{University of the Chinese Academy of Sciences, YuQuanLu 19A, Beijing 100049, China}

\title{Titolo}

\ifthenelse{\equal{\pr}{1}}{
\title{\thetitle}
\author{\autone}
\affiliation {\affa}
\affiliation {\affb}

}{

\begin{center}
{\large {\bf \thetitle}}

\bigskip

\bigskip


{\large \noindent  \autone{${}^{1,2}$} }


\vskip.7cm

1) \affa\\
2) \affb\\

\end{center}

}

\begin{abstract}
\noindent
Recently, Jarah has constructed the kink form factor relevant to the scattering of an ultrarelativistic meson with an arbitrary nonrelativistic scalar kink.  However the formula was only applied to the Sine-Gordon model, where the form factor was long ago determined by Weisz using integrability.  In this paper, using various known analytic results for the (1+1)-dimensional real scalar~$\Phi^4$ model and a kink wave packet construction, we analytically calculate the leading quantum correction to the form factor of the~$\Phi^4$ kink.   We discuss its properties in general and also in the ultra-relativistic meson case and provide a numerical check of our results.

\end{abstract}

%
\setcounter{footnote}{0}
\renewcommand{\thefootnote}{\arabic{footnote}}

\ifthenelse{\equal{\pr}{1}}
{
\maketitle
}{}

\section{Introduction}
The~$\Phi^4$ model describes phenomena in disparate fields and so has long been one of the most studied models in theoretical physics.   The (1+1)-dimensional real scalar $\Phi^4$ model admits a topological soliton called the kink, which describes a stable particle-like field excitation and has many applications in cosmology,  chemistry, particle physics, biology and quantum optics Refs.~\cite{chinesebook, mantonbook,skymion,cosmic string, dark energy, optical soliton, protein1, protein2}. The kink is also of interest because of its simplicity, as the methods developed to study it can in some cases be generalized to (3+1)-dimensional theories like QCD, which is our motivation.\par
The original work connecting soliton\footnote{The kink is a topological soliton in (1+1)-dimensional spacetime.} with particle physics was done by Skyrme in 1960s. He constructed the baryon state out of the pure meson fields as a topological soliton solution now known as the Skyrmion in honor of his creative work in Refs.~\cite{Skyrme:1962vh, Skyrme:1961vq}.  But it did not attract much attention until the 1970s, with the pioneering work of Dashen, Hasslacher  and Neveu who calculated the leading  quantum correction to the mass of the kink of the $\Phi^4$ and Non-Abelian models using a semi-classical method Refs.~\cite{dhn1,dhn2,dhn3}, then followed by the work of  T. D. Lee using a canonical quantization approach on similiar models Ref.~\cite{lee}. \par
Another famous prototype, the Sine-Gordon model Ref.~\cite{sg integrable}, has been discussed widely and deeply. The key difference between the~$\Phi^4$ model and the Sine-Gordon model is the integrability Ref.~\cite{sg integrable}.  While the Sine-Gordon model is integrable, the~$\Phi^4$ model is not.  However, they both possess stable and localized solitary wave configurations of finite energy Ref.~\cite{mantonbook}.  In dynamic collision processes, the kinks in the~$\Phi^4$ model can not pass through one another unaffected as they do in Sine-Gordon model Ref.~\cite{belova}. This particular property results in many interesting and amazing  physical phenomenons like the bounce, resonance, impurity scattering and bion ( breather) formation Ref.~\cite{wiegelcc}. All of these processes have found applications in physics Refs.~\cite{mantonbook,rajaraman,phi4introducntion}.

The key to connecting the model and phenomenolgy is scattering.  It is also the main approach to examine the validity of the model and various methods.  This has attracted physicists to investigate kink dynamics and construct the scatting matrix relevant to the kink, such as the kink-meson scattering and the kink-kink scattering which are analogous to meson baryon scattering, meson absorption and emission by a nucleus and other processes in particle physics Refs.~\cite{weigel2011, hayashi1992}.  Kink scattering is usually treated by using the collective coordinate approach Ref.~\cite{gjs} where the kink position is promoted to an operator. But despite its great success,  it becomes prohibitively complicated at higher orders.\par
Recently a new approach has been proposed \cite{mekink}. The key is the unitary transformation of the original, regularized, defining Hamiltonian to a quantum kink Hamiltonian via a displacement operator which is constructed using the classical kink solution.   The same unitary transformation automatically relates the eigenstates of the two Hamiltonians.  Thus the eigenstates of the kink Hamiltonian, which can be calculated in perturbation theory, can be used to construct the eigenstates of the original, defining Hamiltonian.  Because the two Hamiltonians are related by a similarity transformation, their spectra are equal.  As the defining Hamiltonian is already regularized, usually via normal ordering, one avoids the double regularization of the vacuum sector and the kink sector which was frequently used in previous path integral and canonical quantization schemes.  The awful zero mode problem, which was usually fixed using collective coordinates, is instead fixed using the translation invariance.  Based on this approach, the 2-loop mass and state corrections of the ground kink state and once excited state have been calculated. To avoid difficulties in the relativistic region,  the approach has been limited to the perturbed boosted kink state Refs.~\cite{me2loop,mekink,memovingkink}.  What is more,  they constructed the general kink wave packet based on the kink ground state, then decomposed it into the position representation which can describe the position of the kink. Furthermore, they constructed the general form factor for elastic ultrarelativistic meson scattering off the nonrelativistic kink and also gave the corresponding  exact results for the Sine-Gordon model.\par 
Our goal is to apply the method they developed to the exact real scalar~$\Phi^4$ model in (1+1)-dimensional of spacetime whose analytic linearized perturbation decomposition and subleading state correction were found in Ref.~\cite{2loopphi4}. We begin in Sec.~\ref{revsez} with a review of the linearized soliton sector perturbation and the analytic result for the~$\Phi^4$ model.  In Sec.~\ref{ff}, we review the definition of the kink wave-packet and the general form factor calculation up to the leading and sub-leading correction.  In Sec.~\ref{phi4ff}, we  calculate the leading correction to the form factor for the~$\Phi^4$ model and discuss its 2 different  cases: the general case and the ultrarelativistic meson case. Then we compare the analytical result with the numerical result.   In Sec.~\ref{conclusion}, we give conclusions, describe their limits and the potential for further study.  We also give~\ref{apa} for a numerical result of the general and ultrarelativistic limit result and~\ref{apb} as a comparison with Jarah's result.

\section{Review} \label{revsez}

\subsection{Linearized Kink Perturbation Theory}

Let's consider a general (1+1) dimensional Hamiltonian $H$ with Hamiltonian density $\ch$ given by
\bea
\ch(x)=\frac{1}{2}:(\pi(x))^2:_a+\frac{1}{2}:(\partial_x\phi(x))^2:_a+\frac{1}{\lambda}:V[\sqrt{\lambda}\phi(x)]:_a=\H_0+\H_I. \label{oh}
\eea 
Here $\phi(x)$ is real scalar field in (1+1)-dimensional of spacetime, the~$\pi(x)$ is the conjugate momentum, $\lambda$ is the real coupling constant and $V[\sqrt{\lambda}\phi(x)]$ has 2 degenerate minima with respect to $\phi(x)$. The~$::_a$ means the normal ordering using the plane wave mode expansion operator of $\phi(x)$ and we are working in the Schrodinger picture so operators are time-independent.  

At the classical level, the field operator $\phi(x)$ reduces to a time-dependent function $\phi(x,t)$.  According to the  classical equation of motion and  under nontrivial boundary conditions, the system has a static topological solution called the kink
\beq
\phi(x,t)=f(x) \label{fd}
\eeq   

In the quantum theory, the kink solution and small excitations about it correspond to a space of states called the kink sector, which consists of the kink ground state plus the Fock space of perturbative excitations above it.  Let~$|K\rangle$ be a Hamiltonian eigenstate in the kink sector
\beq
H|K\rangle=E_K|K\rangle 
\eeq 
which satisfies $\langle K|\phi(x)|K\rangle=f(x)$ at leading order in the semiclassical expansion. 
Using the displacement operator
\beq
\df={\rm{exp}}\left(-i\int dx f(x)\pi(x)\right) \label{df}
\eeq
we define the shifted Hamiltonian Ref.~\cite{mekink} 
\beq
H\p=\df^\dag H\df\hsp 
H[\phi,\pi]\rightarrow H\p[\phi,\pi]=H[f+\phi,\pi].\label{act} 
\eeq
Unitary equivalence in~(\ref{act}) means that $H$ and $H\p$ have the same eigenvalues with their eigenvectors related by~$\df$.  This means that we could choose either of them to calculate the eigenvalues.  We will see that perturbation theory may be used to calculate vacuum sector states using~$H$ and kink sector states using~$H\p$.  The 2 Fock spaces are connected by
\beq
|K\rangle=\df|\Omega\rangle
\eeq
where the~$|\Omega\rangle$ is an eigenstate of~$H\p$.  So when $|\Omega\rangle$ is the ground state of the $H\p$, which we denote by $\vac$,  we have the semi-classical expansion of the kink ground state in powers of~$\sqrt{\lambda}\hbar$\footnote{We set $\hbar=1$.} 
\beq
\vac=\sum_{i=0}\vac_i
\eeq
In any case we find the relation:
\beq
H\p|\Omega\rangle=\D_f^{\dagger} H\D_p|\Omega\rangle=\D_f^{\dagger}H|K\rangle=\D_f^{\dagger}E_K|K\rangle=E_K|\Omega\rangle\label{sh}
\eeq
This transformation simplifies the problem and even holds when $|\Omega\rangle=\vac$.   We can of course calculate the all spectrum of the kink under~$H\p$ and subtract the expectation value of vacuum to get the kink mass spectrum.  Therefore we will refer to the shifted Hamiltonian $H\p$ as the kink Hamiltonian.  As~$\sqrt{\lambda}\hbar$ is dimensionless, we expand~$H\p$ in powers of $\sqrt{\lambda}$
\bea
H\p&=&\df^\dag H\df=Q_0+H_2+\sum_{n=2}^{\infty}H_n\hsp\\
H_2&=&\frac{1}{2}\int dx\left[:\pi^2(x):_a+:\left(\partial_x\phi(x)\right)^2:_a+V^{\prime\prime}[\sqrt{\lambda}f(x)]:\phi^2(x):_a\right]\nonumber
\eea
where~$Q_0$ is the classical kink mass and~$V^{(n)}$ is the~$n$-th functional derivative of~$\frac{1}{\lambda}V[g\phi(x)]$ with respect to the~$\phi(x)$.  The classical, linear wave equation corresponding to~$H_2$ is a Sturm-Liouville equation with eigenvalue
\beq
\omega_k=\sqrt{m^2+k^2} \label{ok}
\eeq
where~$\frac{1}{\lambda}V^{\prime\prime}[\lambda\phi(x)]|_{\phi(x)=0}=m^2$ in~(2.1) for the defining Hamiltonian and $m$ here is the mass of the scalar field.  The equation has general solutions~$\g_k(x)$ which consist of continuum  modes $\g_k(x$) with~$\omega_k>m$,  discrete shape modes $\g_k(x)=\g_S(x)$ with~$0<\omega_k<m$ and one zero-mode  $\g_k(x)=\g_B(x)$ with~$\omega_k=0$.  We will often need to integrate over continuum modes and and sum up all shape modes, so we use the symbol~$\int\hspace{-12pt}\sum\frac{dk}{2\pi}$ to include both the integral over continuous modes and sum over the shape modes. It is noted that~$2\pi\delta(k-k\p)$ should be understood as~$\delta_{kk\p}$ when~$k$ represents a shape mode.  Then we impose normalization conditions:
\beq
\int dx \g_{k_1} (x) \g^*_{k_2}(x)=2\pi \delta(k_1-k_2),\ 
\int dx |\g_{B}(x)|^2=1
\eeq
the completeness relation 
\beq
\g_B(x)\g_B(y)+\ppin{k}\g_k(x)\g^*_{k}(y)=\delta(x-y) \label{comp}
\eeq
and convention
\beq
\g_k(-x)=\g_k^*(x)=\g_{-k}(x),\ \tilde{\g}(p)=\int dx \g(x) e^{ipx}
\eeq
What is more, we can do the plane wave expansion
\bea
\phi(x)&=&\ppin{p}\left(A^\dag_p+\frac{A_{-p}}{2\omega_p}\right) e^{-ipx}\label{adec}\\
 \pi(x)&=&i\ppin{p}\left(\omega_pA^\dag_p-\frac{A_{-p}}{2}\right) e^{-ipx}
\nonumber
\eea
and the normal mode expansion Ref.~\cite{me2stato}
\bea
\phi(x)&=&\phi_0 \g_B(x) +\ppin{k}\left(B_k^\dag+\frac{B_{-k}}{2\omega_k}\right) \g_k(x)\\\nonumber
\pi(x)&=&\pi_0 \g_B(x)+i\ppin{k}\left(\omega_kB_k^\dag - \frac{B_{-k}}{2}\right) \g_k(x)\label{bdec}
\eea
where $\{A_{p},A^{\dag}_p\}$ the $\{\Bk{},\Bd{}\}$ are respectively deformed creation and annihilation conjugate operator pairs of plane wave expansion and the normal mode expansion, and the $\phi_0,\pi_0$ indicate the position and momentum operators of the kink center of mass.   It is easy to find the relations:
\bea
[A_p,A_q^\dag]&=&2\pi\delta(p-q)\\
{[\phi_0,\pi_0]}&=&i\hsp
[B_{k_1},B^\dag_{k_2}]=2\pi\delta(k_1-k_2)\nonumber
\eea 

Now consider the kink eigenstate~$|K\rangle$ of $H$ corresponding to the perturbative ground state~$|0\rangle$ of~$H\p$.  We can expand it in powers of~$\sqrt{\lambda}\hbar$
\beq
|0\rangle=\sum_{i=0}^\infty |0\rangle_{i}\hsp  |0\rangle_{i}=\sum_{m,n}|0\rangle_i^{mn}\label{semi}
\eeq
where we defined 
\beq
|0\rangle_i^{mn}
=Q_0^{-i/2}\ppink{n}\gamma_i^{mn}(k_1,k_2,\cdots,k_n)\phi_0^m
B^{\dag}_{k_1}\cdots B^{\dag}_{k_n}|0\rangle_0 \label{gsd}
\eeq 
Here the $\gamma_i^{ij}$ are coefficient functions. The $n$-loop results correspond to truncating at~$i=2n-2$. 
Then we get the diagonalizable Hamiltonian
\bea
H_2&=&Q_1+\frac{\pi_0^2}{2}+\ppin{k}\omega_k B^\dag_k B_k 
\eea
$Q_1$ is the one-loop kink mass.  The ground state~$|0\rangle_0$ of~$H_2$ satisfies
\beq
\pi_0|0\rangle_0=B_k|0\rangle_0=0 \label{v0}
\eeq
which means~$|0\rangle_0$ is the one-loop corrected kink ground state.   To overcome the zero mode problem, we use the translation invariance, which yields the recursion relation Ref.~\cite{mekink}.
\bea
\gamma_{i+1}^{mn}(k_1\cdots k_n)&=&\left.\Delta_{k_n B}\left(\gamma_i^{m,n-1}(k_1\cdots k_{n-1})+\frac{\omega_{k_n}}{m}\gamma_i^{m-2,n-1}(k_1\cdots k_{n-1})\right)\right. \label{rrs}\\
&&+(n+1)\ppin{k\p}\Delta_{-k\p B}\left(\frac{\gamma_i^{m,n+1}(k_1\cdots k_n,k\p)}{2\omega_{k\p}}
-\frac{\gamma_i^{m-2,n+1}(k_1\cdots k_n,k\p)}{2m}\right)\nonumber\\
&&+\frac{\omega_{k_{n-1}}\Delta_{k_{n-1}k_n}}{m}\gamma_i^{m-1,n-2}(k_1\cdots k_{n-2})\nonumber\\
&&+\frac{n}{2m}\ppin{k\p}\Delta_{k_n,-k\p}\left(1+\frac{\omega_{k_n}}{\omega_{k\p}}\right)\gamma^{m-1,n}_i(k_1\cdots k_{n-1},k\p)
\nonumber\\
&&\left.-\frac{(n+2)(n+1)}{2m}\ppinkp{2}\frac{\Delta_{-k\p_1,-k\p_2}}{2\omega_{k\p_2}} \gamma_i^{m-1,n+2}(k_1\cdots k_{n},k\p_1,k\p_2).
\right.
\nonumber
\eea 
The kink ground state corresponds to the initial condition
 \beq
 \gamma_0^{mn}=\delta_{m0}\delta_{no}\gamma_0^{00}.
 \eeq
In this case the recursion yields 
\beas
\gamma_1^{12}(k_1,k_2)&=&\frac{\left(\omega_{k_1}-\omega_{k_2}\right)\Delta_{k_1k_2}}{2}\gamma_0^{00}  \\
\gamma_1^{21}(k_1)&=&\frac{\omega_{k_1}\Delta_{k_1B}}{2}\gamma_0^{00} \label{g121}
\eeas
Then with the Schrodinger equation:
\beq
(H-E)\vac=0
\eeq
at next leading order:
\beq
(H_3-Q_{1.5})|0\rangle_0+(H_2-Q_1)|0\rangle_1=0
\eeq
We can further get the remaining terms with~$\phi_0$ of the~$|0\rangle_1$ exactly.  We write down the terms we need:
\beq
\begin{aligned}
	&\gamma_1^{01}(k_1)=(\frac{\Delta_{k_1B}}{2}-\sqrt{Q_0}\frac{V_{\I k_1}}{\w1})\gamma_0^{00},
	&\gamma_1^{21}(k_1)=\frac{\w1\Delta_{k_1B}}{2}\gamma_0^{00}
\end{aligned}
\eeq 
where we adapt the convention Ref.~\cite{mekink}. 
\beq
\I(x)=\pin{k}\frac{\left|{\g}_{k}(x)\right|^2-1}{2\omega_k}+\sum_S \frac{\left|{\g}_{S}(x)\right|^2}{2\omega_k}.\label{di}
\eeq
\beq
\Delta_{ij}=\int dx \g_i(x)\g_j(x),\hspace{5pt}
 V_{ijl}=\int dx V^{'''}[\sqrt{\lambda}f(x)]\g_i(x)\g_j(x)\g_l(x),\hspace{5pt}
 V_{\I i}=\int dx V^{'''}[\sqrt{\lambda}f(x)]\I(x)\g_i(x) \label{dij}
\eeq
The  indices~$i, j, l$ include the shape, zero and continuous modes.  A useful relation here is
\beq
f\p(x)=\sqrt{Q_0}\g_B(x)
\eeq
This concludes our review of the construction of next leading order kink state in a general model.

\subsection{State correction for the $\Phi^4$ model}\label{phi4r}
For the~$\Phi^4$ model, the potential in~(2.1) is
\beq
 \frac{1}{\lambda}V[\sqrt{\lambda}\phi(x)]=\frac{1}{\lambda}\frac{\lambda\phi^2}{4}(\sqrt{\lambda}\phi-\sqrt{8}{\b})^2
\eeq
where the $\b$,$\lambda$ are both real parameters.  The equation of motion has a classical kink solution
 \beq
 f(x)=\b\sqrt{\frac{2}{\lambda}}(1+\tanh(\b x))
 \eeq
and its normal modes are known analytically
\begin{equation}
\begin{aligned}
&\g_k(x)=\frac{e^{-ikx}}{\omega_k(k^2+\b^2)}[k^2-2\b^2+3\b^2\sech(\b x)-3i\b k\tanh(\b x)]\\
&\g_S(x)=-i\sqrt{\frac{3\b}{2}}\tanh(\b x)\sech(\b x)\hsp \g_B(x)=\frac{\sqrt{3\b}}{2}\sech^2(\b x)\\
&\omega_k=\sqrt{4\b^2+k^2}\hsp \omega_S=\b\sqrt{3}\hsp \omega_B=0 \label{gk}.
\end{aligned}
\end{equation}
 The continuous mode excited state corresponds the kink plus the moving meson of mass $m=2\b$.  Eq~(\ref{dij}) is therefore
\begin{equation}
\begin{aligned}
&\Delta_{SB}=i\pi\frac{3\b}{8\sqrt{2}}\hsp
\Delta_{kB}=i\pi\frac{\sqrt{3}}{8}\frac{k^2\omega_k}{\b^{3/2}\sqrt{\b^2+k^2}}\csch(\frac{\pi k}{2\b})\\
&V_{\I k}=i\frac{\sqrt{\lambda}}
{32\sqrt{6}}\frac{k^2\omega_k}{\b^4\sqrt{\b^2+k^2}}[2\pi(-2\b^2+k^2)+3\sqrt{3}\o{}^2\csch(\frac{\pi k}{2\b})]\hsp
V_{\I B}=i\frac{3\sqrt{\lambda}}{64}\sqrt{\b}(3\sqrt{3}-2\pi)\label{useful}
\end{aligned}
\end{equation}
It has classical kink mass~$Q_0=\frac{8\b^3}{3\lambda}$. These are all the ingredients we need for the analytic calculation of the  form factor of the~$\Phi^4$ model.

\section{Form factor}\label{ff}
In the kink sector, we construct the wave-packet Ref.~\cite{memovingkink}
\beq
\as=\frac{\sqrt{\mathcal{N}}}{(2\pi)^{1/4}\sqrt{\sigma}}e^{-\frac{\phi_0^2}{4\sigma^2}}e^{i\alpha\Lambda\p}\vac \label{sig}
\eeq
Here $\mathcal{N}$ is the normalization constant determined by the normalization condition:
\beq
\langle\alpha;\sigma\as=1 \label{norm}
\eeq
$\alpha$ is the rapidity of the center of mass of the wave packet which will be illustrated later.  $\sigma$ is the width of our wave packet, and $\Lambda\p$ is the shifted boost operator\footnote{We will refer to the shifted boost operator  as the boost operator for simplicity because we focus on the shifted Hamiltonian and state.} defined analogously to $H\p$ (\ref{sh}).  Define the momentum space form factor as
\beq
\tilde{\ff}_{q}=\langle 0;\sigma|\df^{\dag} \tp_q \df \as\label{ffd}
\eeq
while its Fourier transformation to position space is~\cite{gj1975}: 
\beq
\ff(z)=\pin{q} e^{-iqz}\tilde{\ff}_{q}
\eeq
We can see the correction comes from the perturbed ground state~$\vac=\sum_i\vac_i$ and the  boost operator~$\Lambda\p=\sum_i \Lambda\p_i$ Refs.~\cite{memovingkink,me2loop}, where the leading and subdominant  corrections to the boost and the kink ground state are given Refs.~\cite{meff, memovingkink} 
\beq
\as_0=\frac{\sqrt{\mathcal{N}}}{(2\pi)^{1/4}\sqrt{\sigma}}e^{-\frac{\phi_0^2}{4\sigma^2}}e^{i\alpha\Lambda_1\p}\vac_0,\hspace{5pt}
\as_{0,1}=\frac{1}{(2\pi)^{1/4}\sqrt{\sigma}}e^{-\frac{\phi_0^2}{4\sigma^2}}e^{i\alpha\Lambda\p_1}\vac_1\label{sig}
\eeq
\beq
\as_{1,0}+\as_0=\frac{\sqrt{\mathcal{N}}}{(2\pi)^{1/4}\sqrt{\sigma}}e^{-\frac{\phi_0^2}{4\sigma^2}}e^{i\alpha(\Lambda_1\p+\Lambda_2\p)}\vac_0
\eeq
\beq
{}_{0,1}\langle 0;\sigma|=\frac{1}{(2\pi)^{1/4}\sqrt{\sigma}}\ {}_1\langle 0|e^{-\frac{\phi_0^2}{4\sigma^2}}\hsp
{}_{0}\langle 0;\sigma|=\frac{1}{(2\pi)^{1/4}\sqrt{\sigma}}\ {}_0\langle 0|e^{-\frac{\phi_0^2}{4\sigma^2}}
\eeq
\beq
\Lambda\p_1=-\sqrt{Q_0}\phi_0
\eeq
\beq
\Lambda\p_2=\ppink{2}\frac{\Delta_{k_1k_2}}{\omega_{k_2}^2-\omega_{k_1}^2}:\left(\pi_{k_1}\pi_{k_2}+\omega_{k_1}^2\phi_{k_1}\phi_{k_2}\right):_b+\ppin{k}\Delta_{B k}\left(\frac{2}{\omega^2_k}\pi_{0}\pi_{k}+ \phi_{0}\phi_{k}\right)  \label{l2}
\eeq
In the calculation of the form factors, it is convenient to introduce the states $|y\rangle_0$ which satisfy
\beq
\vac_0=\int dy |y\rangle_0\hsp
\phi_0|y\rangle_0=y|y\rangle_0\hsp B_k|y\rangle_0=0.
\eeq
These play a role similar to position eigenstates for the kink.

\subsection{Form factor at tree level and leading correction}
With above exact corrections of the boost and ground kink state, we review the tree level and leading order  form factor.  The tree level form factor is:
\bea
\tilde{\ff}_{{\rm{tree}},q}&=&{}_0\langle 0;\sigma|\df^\dag \tilde{\phi_q}\df\as_0
 =\int dx e^{iqx}{}_0\langle 0;\sigma|\df^\dag \phi(x) \df\as_0\label{ff00}\\\nonumber
&=&\frac{1}{\sigma\sqrt{2\pi}}\int dx e^{iqx}\int dy e^{-\frac{y^2}{2\sigma^2}-i\alpha\sqrt{Q_0}y}\left(f(x)+\frac{y}{\sqrt{Q_0}}f\p(x)
\right)
\eea
Then with the variable transformation~$z=x+\frac{y}{\sqrt{Q_0}}$, where~$\frac{y}{\sqrt{Q_0}}$ indicates the center of mass of the kink, we can get the leading term in the tree level form factor:
\beq
\tilde{\ff}_{0,q}=
\frac{1}{\sigma\sqrt{2\pi}}\int dz e^{iqz}f(z)\int dy e^{-\frac{y^2}{2\sigma^2}-i(Q_0\alpha+q)y/\sqrt{Q_0}}=\int dz e^{iqz}f(z)e^{-\frac{\sigma^2\left(Q_0\alpha+q\right)^2}{2Q_0}}
\eeq
We find that the wave packet center carries momentum~$Q_0\alpha$ and the form factor distribution coefficient in momentum space is peaked at~$q=-Q_0\alpha$ and dominant near the peak
Ref.~\cite{memovingkink}.  Also, for finite~$\sigma$, it has momentum spread which is of order~$\sqrt{Q_0}/\sigma$.  Thus a nontrivial contribution happens only when~$q$ closes to~$-Q_0\alpha$ and the form factor amplitude is largest at~$q=-Q_0\alpha$.\par 
If we set:~$q=-Q_0\a$ in the localized limit, we have 
\beq
\tilde{\ff}_{0,q=-Q_0\alpha}=\int dz e^{iqz}f(z)
\eeq
And in a general case, we  can set:~$q=\epsilon-Q_0\alpha$.
\beq
\tilde{\ff}_{0,q=\epsilon-Q_0\alpha}=e^{-\frac{\sigma^2\epsilon^2}{2Q_0}}\int dz e^{iqz}f(z)\hsp 
\ff_{0,\epsilon}(z)=\pin{q} e^{-iqz}\tilde{\ff}_{0,q=\epsilon-Q_0\alpha}=e^{-\frac{\sigma^2\epsilon^2}{2Q_0}}f(z) \label{ff0}
\eeq
Next we come to the second order derivative term in~(\ref{ff00})\footnote{This contributes the next leading correction in the Taylor expansion of $f(x)$ in Eq.~(\ref{ff00}).}
\bea
\tilde{\cc}_{1,q}&=&-\frac{1}{2}
\int dz e^{iqz}\int dy \frac{e^{-\frac{y^2}{2\sigma^2}-i(Q_0\alpha+q)y/\sqrt{Q_0}}}{\sigma\sqrt{2\pi}}\left(\frac{y}{\sqrt{Q_0}}\right)^2  f\pp(z)\\
&=&\int dz e^{iqz}\left[
-\frac{f\pp(z)}{2Q_0}\sigma^2\left(1-\frac{\sigma^2\left(Q_0\alpha+q\right)^2}{Q_0}\right)e^{-\frac{\sigma^2\left(Q_0\alpha+q\right)^2}{2Q_0}}
\right]\nonumber
\eea
which contributes: 
\beq
\cc_{1,\epsilon}(z)=-\frac{f\pp(z)}{2Q_0}\sigma^2\left(1-\frac{\sigma^2\epsilon^2}{Q_0}\right)e^{-\frac{\sigma^2\epsilon^2}{2Q_0}} \hsp
\cc_{1,\epsilon=0}(z)=-\frac{f\pp(z)}{2Q_0}\sigma^2  \label{c10}
\eeq
It is of order
\beq
\cc_{1}(z)\sim O(g(\sigma^2m))
\eeq
Here we denote~$g=\sqrt{\lambda}$ as perturbation parameter for sake of comparing with the convention in Ref.~\cite{meff} and (2.1).
\par

The leading correction to the boost operator part is 
\bea
\tilde{\cc}_{2,q}&=&{}_0\langle 0;\sigma|\df^\dag \tp_q \df \as_{1,0}={}_0\langle 0;\sigma| \tp_q  \as_{1,0}\\
&=&\int dx e^{iqx}\frac{i\alpha}{\sigma\sqrt{2\pi}}\left[\ppin{k} \frac{\g_{-k}(x)\Delta_{k B}}{2\ok{}}\right]\int dy y e^{-\frac{y^2}{2\sigma^2}-i\alpha\sqrt{Q_0}y}
\nonumber
\eea
It contributes at order
\beq
\cc_{2,\epsilon}(z)\sim O(g^3(\sigma^2m))
\eeq
It is a subleading correction, so we don't need to consider it.\par

The last contribution of order~$O(g(\sigma^2m))$ comes from:
 \bea
 \tilde{\cc}_{3,q}&=&{}_0\langle 0;\sigma|\df^\dag \tp_q \df \as_{0,1}+{}_{0,1}\langle 0;\sigma|\df^\dag \tp_q \df \as_{0}\\\nonumber
 &=&\int dz e^{iqz} 
 \bigg[\frac{2}{\sqrt{Q_0}}\frac{1}{\sigma\sqrt{2\pi}}\int dy \left[\ppin{k} \frac{\g_{-k}\left(z-\frac{y}{\sqrt{Q_0}}\right)}{2\ok{}}\right] \\\nonumber
 &&\times \left[{\gamma_1^{01}(k)}+y^2{\gamma_1^{21}(k)}\right] e^{-\frac{y^2}{2\sigma^2}-i(Q_0\alpha+q)y/\sqrt{Q_0}}
 \bigg]\nonumber
 \eea
With similiar manipulations, we get its  position coordinate representation: 
\beq
\cc_{3,\epsilon}(z)=\frac{2}{\sqrt{Q_0}}e^{-\frac{\sigma^2\epsilon^2}{2Q_0}}
\ppin{k} \frac{\g_{-k}\left(z\right)}{2\ok{}} \left[{\gamma_1^{01}(k)}+\sigma^2\left(1-\frac{\sigma^2\epsilon^2}{Q_0}\right){\gamma_1^{21}(k)}
\right] \label{c30}
\eeq
Reviewing the next leading order state correction in Sec.~\ref{revsez}
  \beq
\gamma_1^{21}(k)=\frac{\ok{}\Delta_{kB}}{2}\hsp
\gamma_1^{01}(k)=\frac{\Delta_{kB}}{2}-\frac{g\sqrt{Q_0}}{2\ok{}}\int dx \V3 \I(x) \g_k(x) \label{gam}
  \eeq
and the integral:
\beas
\ppin{k} \frac{\g_{-k}(z)}{2\ok{}}\gamma_1^{21}(k)&=&\frac{1}{4}\int dx \ppin{k} \g_{-k}(z) \g_k(x) \g\p_B(x)\\\nonumber
&=&\frac{1}{4}\int dx \left(\delta(x-z)-\g_B(x) \g_B(z)  \right)\g\p_B(x)\\\nonumber
&=&\frac{\g\p_B(z)}{4}=\frac{f\pp_B(z)}{4\sqrt{Q_0}}
\eeas
which cancels the~$\cc_{1,\epsilon}$. We conclude that the total contribution to the form factor correction at order~$O(g(\sigma^2m))$ is
\bea
\cc_{1,\epsilon}(z)&+&\cc_{3,\epsilon}(z)=\frac{2}{\sqrt{Q_0}}e^{-\frac{\sigma^2\epsilon^2}{2Q_0}}
\ppin{k} \frac{\g_{-k}\left(z\right)}{2\ok{}}{\gamma_1^{01}(k)}\label{c3}\\
&=&\frac{2}{\sqrt{Q_0}}e^{-\frac{\sigma^2\epsilon^2}{2Q_0}}
\ppin{k} \frac{\g_{-k}\left(z\right)}{2\ok{}} \left[\frac{\Delta_{kB}}{2}-\frac{g\sqrt{Q_0}}{2\ok{}}V_{\I k}
\right]\label{gorder}
\eea
We finished the review of the leading order correction to the form factor.  We are now ready to calculate the exact result for the~$\Phi^4$ model, which is our main goal.

\section{Form factor of the $\Phi^4$ model }\label{phi4ff}
\subsection{Main Result}
In the case of the $\Phi^4$ kink, which has a single shape mode, the general formula (\ref{gorder}) can be decomposed
\begin{equation}
	\begin{aligned}
\cc_{1}(z)+\cc_{3}(z)
=&\frac{1}{\sqrt{Q_0}}e^{-\frac{\sigma^2\epsilon^2}{2Q_0}}
\pin{k}\frac{\g_{-k}\left(z\right)}{2\ok{}} \left[\Delta_{kB}-\frac{\sqrt{Q_0}}{\ok{}}V_{\I k}\right]\\
 &+\frac{1}{\sqrt{Q_0}}e^{-\frac{\sigma^2\epsilon^2}{2Q_0}}\frac{\g_{-S}\left(z\right)}{2\os{}} \left[\Delta_{SB}-\frac{\sqrt{Q_0}}{
 	\ok{}}V_{\I S}\right]
=\frac{1}{\sqrt{Q_0}}e^{-\frac{\sigma^2\epsilon^2}{2Q_0}}(A+B)
 \end{aligned}
\end{equation}
We substitute the exact results (\ref{gk}) and (\ref{useful}) for the $\Phi^4$ model into the above, using
\beq
\lambda=g^2
\eeq
to obtain
\begin{equation}
	\begin{aligned}
		A=&\pin{k}\frac{\g_{-k}\left(z\right)}{2\ok{}} \left[\Delta_{kB}-\frac{\sqrt{Q_0}}{\ok{}}V_{\I k}\right]\\
		=&\pin{k}\frac{e^{ikz}}{2\o{}^2\sqrt{k^2+\b^2}}[k^2-2\b^2+3\b^2\sech(\b z)+3i\b k\tanh(\b z)]\\
		&\times\left[i\pi\frac{\sqrt{3}}{8}\frac{k^2\omega_k\csch(\frac{\pi k}{2\b})}{\b^{3/2}\sqrt{\b^2+k^2}}
		-\frac{i\sqrt{Q_0\lambda}k^2\omega_k}{32\sqrt{6}\ok{}\b^4\sqrt{\b^2+k^2}}[2\pi(-2\b^2+k^2)+3\sqrt{3}\o^2
		]\csch(\frac{\pi k}{2\b})\right]\\
		=&i\pin{k}e^{ikz}\csch(\frac{\pi k}{2\b})\bigg[\frac{\pi\sqrt{3}}{16\b^{-1/2}\o{}}
		-\frac{(2\pi+3\sqrt{3}-12\pi\frac{\b^2}{\o^2})}{96\b^{1/2}}\bigg]\frac{\frac{k^2}{\b^2}+3i\frac{k}{\b}\tanh(\b z)-2+3\sech(\b z)}{\b^2/k^2+1}\\
		=& \frac{i}{\b^{1/2}}\pin{k}e^{ikz}\csch(\frac{\pi k}{2\b})\bigg[\frac{\pi\sqrt{3}}{16\sqrt{4+\frac{k^2}{\b^2}}}
		-\frac{(2\pi+3\sqrt{3}-12\pi\frac{\b^2}{\o^2})}{96}\bigg]\frac{\frac{k^2}{\b^2}+3i\frac{k}{\b}\tanh(\b z)-2+3\sech(\b z)}{\b^2/k^2+1}
	\end{aligned}
\end{equation}
and
\begin{equation}
	\begin{aligned}
	B=&\frac{\g_{-S}\left(z\right)}{2\os{}} \left[\Delta_{SB}-\frac{\sqrt{Q_0}}{\omega_S{}}V_{\I S}\right]	\\
	=&\frac{-i\sqrt{\frac{3\b}{2}}}{2\omega_S}\tanh(\b x)\sech(\b x)
	\times\left[\i\pi\b\frac{3\pi}{8\sqrt{2}} -\frac{\sqrt{Q_0}}{\omega_S}i\frac{3\sqrt{\lambda\b}}{64}(3\sqrt{3}-2\pi)\right]\\
	=&\sqrt{\b}\frac{12\sqrt{3}\pi^2-3\sqrt{6}+2\sqrt{2}\pi}{64\sqrt{3}}\tanh(\b z)\sech(\b z)
	=\sqrt{\b}(\frac{3}{16}\pi^2-\frac{3\sqrt{2}}{64}+\frac{2\sqrt{6}{\pi}}{384})\tanh(\b z)\sech(\b z)	
	\end{aligned}
\end{equation}
So now we have the general result of the form factor at order~$O(g(\sigma^2m))$.

Recall that the meson momentum is described by a Gaussian wave function.  We will now fix $\epsilon=0$, which means that the peak of that wave function has the same momentum as the kink recoil, so that momentum conservation chooses the peak of the wave function and the form factor is maximized.  Then at leading order we drop the $\sigma$ exponential term\footnote{Which is equal 1 when $\epsilon=0$.} to get:
\begin{equation}
	\begin{aligned}
&\cc_{1}(z)+\cc_{3}(z)=\frac{1}{\sqrt{Q_0}}(A+B)\\
=&\frac{1}{\sqrt{\frac{8\b^3}{3\lambda}}}\bigg[ \frac{i}{\b^{1/2}}\pin{k}e^{ikz}\csch(\frac{\pi k}{2\beta})\bigg[\frac{\pi\sqrt{3}}{16\sqrt{4+\frac{k^2}{\b^2}}}
-\frac{(2\pi+3\sqrt{3}-12\pi\frac{\b^2}{\o^2})}{96}\bigg]\\
&\times\frac{\frac{k^2}{\b^2}+3i\frac{k}{\b}\tanh(\b z)-2+3\sech(\b z)}{\b^2/k^2+1}+\sqrt{\b}(\frac{3}{16}\pi^2-\frac{3\sqrt{2}}{64}+\frac{2\sqrt{6}{\pi}}{384})\tanh(\b z)\sech(\b z)\bigg]\\
=&i\sqrt{\lambda}\bigg[ \frac{1}{\b^2}\pin{k}e^{ikz}\csch(\frac{\pi k}{2\beta})\bigg[\frac{3\pi}{32\sqrt{8+2\frac{k^2}{\b^2}}}
-\frac{\sqrt{3}(2\pi+3\sqrt{3}-12\pi\frac{\b^2}{\o^2})}{192\sqrt{2}}\bigg]\\
&\times\frac{\frac{k^2}{\b^2}+3i\frac{k}{\b}\tanh(\b z)-2+3\sech(\b z)}{\b^2/k^2+1}+\frac{1}{\b}(\frac{3\sqrt{3}}{32\sqrt{2}}\pi^2-\frac{3\sqrt{3}}{128}+\frac{3\pi}{384})
\tanh(\b z)\sech(\b z)\bigg]   \label{gff0}
  \end{aligned}
\end{equation}
In order to make the boost correction converge, we need  the expansion parameter~$\a^2/g\ll 1$.  This means for example that the kink kinetic energy which is of order ~$Q\a^2 \sim m\a^2/g$ should be less than~$Qg\sim mg$, so the kinetic energy may be greater than the meson mass.  In a word,  the result is only valid in the ultrarelativistic meson regime with the fact~$k\gg m=2\b$.  And it is noted that:~$\sech(x),\tanh(x)\in(0,1)$ in~(\ref{gff0}), so we can simplify the~$\cc_{1}(z)+\cc_{3}(z)$ by dropping the subdominant terms above and keeping only the part from term A: 
\beq
\begin{aligned}
\cc_{1}(z)+\cc_{3}(z)
 \approx&-i\sqrt{\lambda}\frac{\sqrt{3}(2\pi+3\sqrt{3})}{192\sqrt{2}\b^2}\pin{k}e^{ikz}\csch(\frac{\pi k}{2\b})\frac{k^2}{\b^2} \\
 =&-i\sqrt{\lambda}\frac{\sqrt{3}(2\pi+3\sqrt{3})}{192\sqrt{2}\b^4}\frac{1}{\sqrt{2\pi}}\times[2i\b^3\sqrt{\frac{2}{\pi}}\sech(\b z)^2\tanh(\b z)]\\
 =&\frac{\sqrt{3\lambda}(2\pi+3\sqrt{3})}{96\sqrt{2}\pi}\sech(\b z)^2\tanh(\b z)		 \label{gff}
\end{aligned}
\eeq
Using classical solution for the $\Phi^4$ model in Subsection.~\ref{phi4r}
 \beq
 f(x)=\b\sqrt{\frac{2}{\lambda}}(1+\tanh(\b x))\hsp
 f^{''}(x)=-\sqrt{\frac{2}{\lambda}}\b^3\sech(\b x)^2\tanh(\b x)
 \eeq
we obtain our main result.  The form factor relevant to kink-meson scattering in the~$\Phi^4$ model in the ultrarelativistic~$k\gg m$ meson limit is:
\beq
\begin{aligned}
\cc_{1}(z)+\cc_{3}(z)=& -\frac{\sqrt{3\lambda}(2\pi+3\sqrt{3})}{96\pi\b\sqrt{2}}\times\frac{1}{\sqrt{\frac{2}{\lambda}}\b^3}f^{''}(z)=-\frac{\sqrt{3}\lambda(2\pi+3\sqrt{3})}{192\pi\b^4}f^{''}(z)\\
=&-0.032963\frac{\lambda}{\b^4}f''(z)= -0.527408\frac{\lambda}{m^4}f''(z)\label{nrlimit}
\end{aligned}
\eeq
This is consistent with unpublished results obtained by Zoltan Bajnok and Marton Lajer obtained using Hamiltonian truncation~\cite{private}.  It indicates that the next order correction of the  form factor for ultrarelativistic meson emission and absorption by a kink  is proportional to its second derivative with perturbatively small and negative coefficients compared with the mass and original coupling constant. This is also a consistency check that we can identify the $\sigma=0$ term in the form factor of localized kink with those of the delocalized kinks Ref.~\cite{memovingkink}.\par
The denominator~$m^4$ seems different from the Sine-Gordon model in appendix B, which is just~$m^2$ once we substitute~$Q_0$ with $m$ in Ref.~\cite{meff}.  This is because $\lambda/m^2$ in the $\Phi^4$ model is dimensionless, playing the role of $g$ in the Sine-Gordon model.
We can therefore see the 2 cases are consistent at the level of dimensional analysis. \par

\subsection{Comparison with Numerical Result}
This leading correction to the Fourier transformed form factor is evaluated numerically in~\ref{apa} and is plotted in Figure~1.\par 
\begin{figure}
	\begin{minipage}[t]{0.4\linewidth}
		\centering
		\includegraphics[width=2.7in]{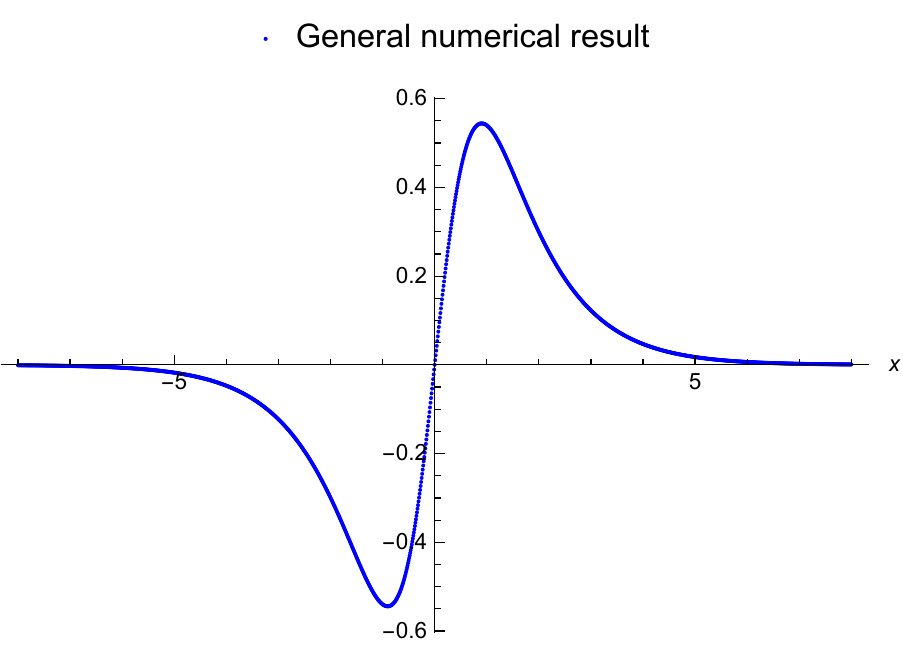}
	\end{minipage}
	\begin{minipage}[t]{0.8\linewidth}
		\centering
		\includegraphics[width=4.1in]{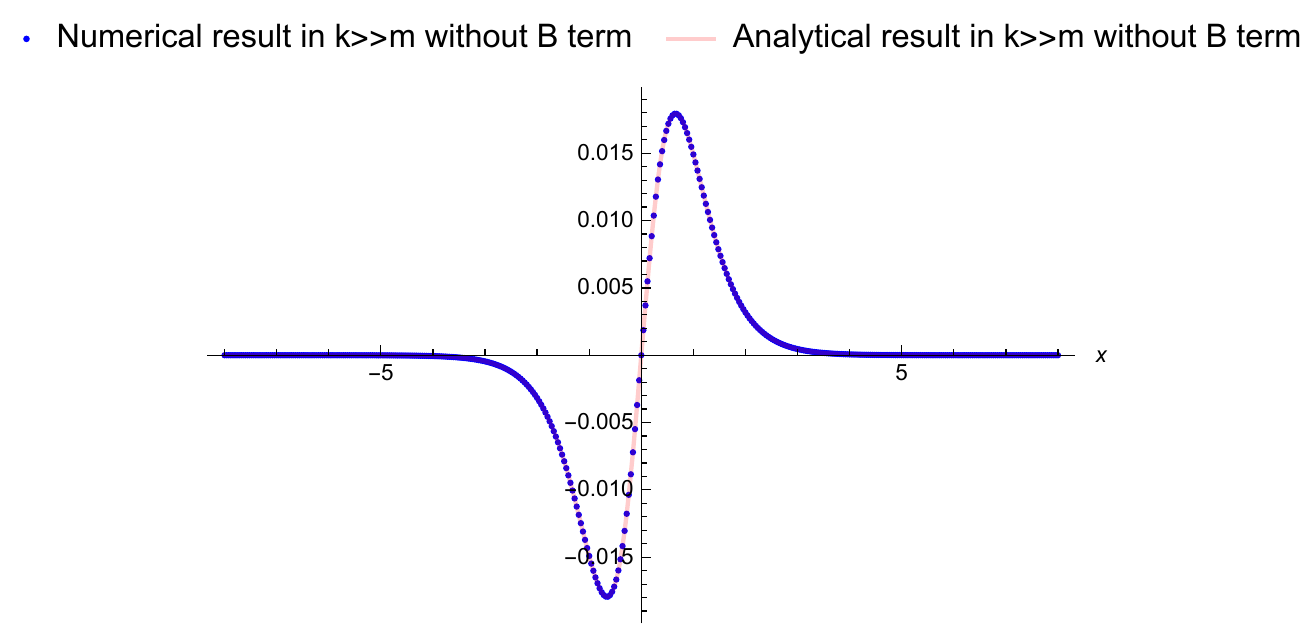}
	\end{minipage}
\caption{We fix our units such that the meson mass is~$m=2\b=2$.  In the diagram, the first panel is the general form factor result from (\ref{gff0}) with the integration performed numerically without any ultrarelativistic approximation.  In the second panel, the dotted blue curve is the form factor of~(\ref{numr}) in the $k\gg m$ limit (ultrarelativistic meson limit) where we used the approximation condition~$k\gg m$ after the the first equality of~(\ref{numr}) and abandoned the shape mode terms which is exactly term B.    It agrees well with -0.032963$f^{''}(x)$ in (\ref{nrlimit}) which is the red solid line in the second panel. 
}
\end{figure}

We can see that the general leading correction (\ref{gff0}) to the form factor has a much bigger amplitude than the ultrarelativistic case $k\gg m$ (\ref{gff}) but a similar shape.  This is because the main contribution of the total form factor correction in (\ref{numr})  comes from the shape mode part~(the $B$ term in (\ref{gff0})).   We can explain this as follows.

The Fourier transformed form factor can be decomposed as
\beq
\Phi(x)=f(x)+a_0\g_S+a_1\g_B+\int\frac{dk}{2\pi}a_k\g_k(x) \label{fexpansion}
\eeq
The coefficients are in general undetermined but are constrained by the fact that the kink form factor at leading order must be the classical solution~$f(x)$ while the excited states contribute the quantum corrections.  With the analysis of the different mode corrections on the classical solution in~(\ref{fexpansion}), we find that the shape mode contribution deforms the classical solution in a direction vertical to the moving direction (that is why it is called a shape mode), the zero mode just has the effect of translating the classical solution.  Although it is still not clear how to choose the perturbation coefficient exactly from general arguments, our numerical result shows that the shape mode has much bigger correction to the form factor compared with the zero mode with same perturbative order.  This agrees with our numerical result in Figure~1.  Of course,  the exact quantitative analysis is waiting for further research.  Our current perturbative expansion is not reliable beyond the ultrarelativistic meson approximation.\par 
It has long been known that kink-antikink scattering in the relativistic region, as can be seen using the collective coordinate method and the numerical simulations, has a rich phenomenology including normal undeformed asymptotic structure, bions and resonances Refs.~\cite{phi4introducntion,Alonso-Izquierdo:2021ygs}.  In such studies a central role is played by the construction of the incident wave function which includes the classical kink and antikink functions and also the shape mode functions.  The initial wave function is characterized by 2 important parameters\footnote{We do not mention the center of mass.}, the velocity of the initial kink and the amplitude before the shape mode function. The initial velocity characterizes the relativistic property of the kink, its effect is well-known to be important and we don't discuss here because we focus on the nonrelativistic kink in our case.   The coefficient before the shape mode function is what we want to connect with our case because it is important  for the rich  structure of asymmetric scattering between kinks and wobblers\footnote{Classical solution plus shape mode function, and it is sometimes called kinks in some literatures without differentiation and so the kink-antikink scattering I mention above indicates the scattering between kinks and wobblers(or the wobller-antiwobbler scattering).} Refs.~\cite{Alonso-Izquierdo:2021ygs}.  When we just abandon the shape mode contribution in our form factor calculation, we indeed ignore its role in scattering between kinks and wobblers(or the antiwobbler-wobbler scattering) we mentioned above.  This illustrates why our analytical result is rather trivial despite the intricate structure which has been revealed in the collective coordinate method.


\section{Conclusion}\label{conclusion}
In conclusion: Our analytical perturbation method and result for the form factor are only valid for ultrarelativistic mesons~($k\gg m$).  As a result, in the integral with abandoned of shape mode contribution as we showed in (4.8) and it matches quantitatively with other methods in this case.  The numerical result shows that for the general Fourier transformed form factor,  most of the contribution comes from the meson without the $k\gg m$ condition part instead of the ultrarelativistic meson, where it is not reliable.   In other words, the Fourier transformed form factor is dominated by the region in momentum space where our perturbative expansion cannot be trusted, and so only the momentum space form factor, at $k\gg m$, should be trusted.

Combined with the  basic  definition of the form factor~(\ref{ffd}) and from the view of physical scattering where the kink represents the nucleon in (1+1) dimensional space, our perturbation method for the form factor calculation is only valid for hard meson scattering with a nonrelativistic nucleon\footnote{Low speed and perturbed moving nucleon.}.  This is the limitation of our method, but with the numerical result from it and the comparison, we indeed can glimpse the fruitful  properties of kink-meson scattering structure just like the rich and astonishing phenomena that arise in kink-antikink scattering with different initial velocity  which need a careful numerical treatment even with the collective coordinates method Ref.~\cite{wiegelcc}. It needs further study both on the collective coordinate approach and even with the improved method we adapt.

\appendix
\section{Numerical calculation for general form  factor}\label{apa}
We begin with the no approximation result~(\ref{gff0})  of form factor and using  the variable replacement~$q=k/\b,x=\b z$ which are dimensionless to get: 
\beq
\begin{aligned}
 \cc_{1}(x)+\cc_{3}(x)=&i\sqrt{\lambda}\bigg[ \frac{1}{\b^2}\pin{k}e^{ikz}\csch(\frac{\pi k}{2\beta})\bigg[\frac{3\pi}{32\sqrt{8+2\frac{k^2}{\b^2}}}
 -\frac{\sqrt{3}(2\pi+3\sqrt{3}-12\pi\frac{\b^2}{\o^2})}{192\sqrt{2}}\bigg]\\
 &\times\frac{\frac{k^2}{\b^2}+3i\frac{k}{\b}\tanh(\b z)-2+3\sech(\b z)}{\b^2/k^2+1}+\frac{1}{\b}(\frac{3\sqrt{3}}{32\sqrt{2}}\pi^2-\frac{3\sqrt{3}}{128}+\frac{3\pi}{384})
 \tanh(\b z)\sech(\b z)\bigg]   \\
 =&i\sqrt{\lambda}\bigg[ \frac{1}{\b}\int\frac{dq}{2\pi}e^{iq x}\csch(\frac{\pi q}{2})\bigg[\frac{3\pi}{32\sqrt{8+2q^2}}
 -\frac{\sqrt{3}(2\pi+3\sqrt{3}-\frac{12\pi}{4+q^2})}{192\sqrt{2}}\bigg]\\
 &\times\frac{q^2+3iq\tanh(x)-2+3\sech(x)}{1/q^2+1}+\frac{1}{\b}(\frac{3\sqrt{3}}{32\sqrt{2}}\pi^2-\frac{3\sqrt{3}}{128}+\frac{3\pi}{384})
 \tanh(x)\sech(x)\bigg]  \\
 =&i\sqrt{\lambda}\bigg[\frac{3\pi}{32\b}\int\frac{dq}{2\pi}e^{iq x}\csch(\frac{\pi q}{2})\frac{q^2+3iq\tanh(x)-2+3\sech(x)}{(1/q^2+1)\sqrt{8+2q^2}}\\
 &-\frac{\sqrt{3}(2\pi+3\sqrt{3})}{192\sqrt{2}\b}\int\frac{dq}{2\pi}e^{iq x}\csch(\frac{\pi q}{2})\frac{q^2+3iq\tanh(x)-2+3\sech(x)}{(1/q^2+1)}\\
 &+\frac{\sqrt{3}\pi}{16\sqrt{2}\b}\int\frac{dq}{2\pi}e^{iq x}\csch(\frac{\pi q}{2})\frac{q^2+3iq\tanh(x)-2+3\sech(x)}{(1/q^2+1)(4+q^2)}\\
 &+\frac{1}{\b}(\frac{3\sqrt{3}}{32\sqrt{2}}\pi^2-\frac{3\sqrt{3}}{128}+\frac{3\pi}{384})
 \tanh(x)\sech(x)\bigg]  \\
 =&\frac{1}{\sqrt{\lambda}\b}\bigg[\frac{3\pi}{32\b}\int\frac{dq}{2\pi}\csch(\frac{\pi q}{2})\frac{-(q^2-2+3\sech(x))\sin(qx)-3q\tanh(x)\cos(qx)}{(1/q^2+1)\sqrt{8+2q^2}}\\
 &-\frac{\sqrt{3}(2\pi+3\sqrt{3})}{192\sqrt{2}\b}\int\frac{dq}{2\pi}\csch(\frac{\pi q}{2})\frac{-(q^2-2+3\sech(x))\sin(qx)-3q\tanh(x)\cos(qx)}{(1/q^2+1)}\\
 &+\frac{\sqrt{3}\pi}{16\sqrt{2}\b}\int\frac{dq}{2\pi}\csch(\frac{\pi q}{2})\frac{-(q^2-2+3\sech(x))\sin(qx)-3q\tanh(x)\cos(qx)}{(1/q^2+1)(4+q^2)}\\
 &+\frac{1}{\b}(\frac{3\sqrt{3}}{32\sqrt{2}}\pi^2-\frac{3\sqrt{3}}{128}+\frac{3\pi}{384})
 \tanh(x)\sech(x)\bigg] \label{numr}
\end{aligned}
\eeq
We can get its result in the general case and the ultrarelativistic meson case~$k\gg \b$ (or $m$) of (4.6) and plot them respectively and also the result of~(\ref{nrlimit}) in Figure~1.\par

\section{Comparison with Jarah and Weisz result for Sine-Gordon}\label{apb}
With the  kinetic mass term result~$Q_0=\frac{8m}{g^2}$ and the form factor correction~(6.7) in the Jarah's paper Ref.~\cite{meff}\footnote{Weisz's result is consistent with Jarah's result which is given as a comparison in Jarah's paper Ref.~\cite{meff} too.}:
\beq
\cc_{1}(z)+\cc_{3}(z)=-\frac{-4f^{''}(z)}{\pi g^2Q_0^2}=-\frac{-4f^{''}(z)}{\pi g^2(\frac{8m}{g^2})^2}=-\frac{g^2f^{''}(z)}{16\pi m^2}=-0.0198\frac{g^2}{m^2}f^{''}(z)
\eeq
Compared with our~(\ref{nrlimit}), we can see that for both the~$\Phi^4$ and Sine-Gordon model, the next leading order correction of form factor in ultrarelativistic meson region is a small correction coefficient  with the second order derivation of the classical solution, this is a consistency check of the perturbative description of the wave packet and state expansion.

\section* {Acknowledgement}

\noindent
The author thanks the Jarah Evslin for  kind supervision and crucial discussion, I also appreciate Zoltan Bajnok and Marton Lajer  for kindly giving their unpublished result with truncated Hamiltonian method to have a comparison. The author also thanks the support of Institute of Modern Physics of CAS. The supervisor of the author  Jarah Evslin and the author is supported by the CAS Key Research Program of Frontier Sciences grant QYZDY-SSW-SLH006 and the NSFC MianShang grants 11875296 and 11675223.   JE also thanks the Recruitment Program of High-end Foreign Experts for support.

\end{document}